\overfullrule = 0pt

\font\bigg=cmbx10 at 17.3 truept    \font\bgg=cmbx10 at 12 truept
\font\twelverm=cmr10 scaled 1200    \font\twelvei=cmmi10 scaled 1200
\font\twelvesy=cmsy10 scaled 1200   \font\twelveex=cmex10 scaled 1200
\font\twelvebf=cmbx10 scaled 1200   \font\twelvesl=cmsl10 scaled 1200
\font\twelvett=cmtt10 scaled 1200   \font\twelveit=cmti10 scaled 1200
\def\twelvepoint{\normalbaselineskip=12.4pt
  \abovedisplayskip 12.4pt plus 3pt minus 9pt
  \belowdisplayskip 12.4pt plus 3pt minus 9pt
  \abovedisplayshortskip 0pt plus 3pt
  \belowdisplayshortskip 7.2pt plus 3pt minus 4pt
  \smallskipamount=3.6pt plus1.2pt minus1.2pt
  \medskipamount=7.2pt plus2.4pt minus2.4pt
  \bigskipamount=14.4pt plus4.8pt minus4.8pt
  \def\rm{\fam0\twelverm}          \def\it{\fam\itfam\twelveit}
  \def\sl{\fam\slfam\twelvesl}     \def\bf{\fam\bffam\twelvebf}
  \def\mit{\fam 1}                 \def\cal{\fam 2}
  \def\tt{\twelvett}
  \textfont0=\twelverm   \scriptfont0=\tenrm   \scriptscriptfont0=\sevenrm
  \textfont1=\twelvei    \scriptfont1=\teni    \scriptscriptfont1=\seveni
  \textfont2=\twelvesy   \scriptfont2=\tensy   \scriptscriptfont2=\sevensy
  \textfont3=\twelveex   \scriptfont3=\twelveex  \scriptscriptfont3=\twelveex
  \textfont\itfam=\twelveit
  \textfont\slfam=\twelvesl
  \textfont\bffam=\twelvebf \scriptfont\bffam=\tenbf
  \scriptscriptfont\bffam=\sevenbf
  \normalbaselines\rm}
  %for real numbers
  %for integer numbers
 %for complex numbers 
\def\IZ{\relax\ifmmode\mathchoice
{\hbox{\cmss Z\kern-.4em Z}}{\hbox{\cmss Z\kern-.4em Z}}
{\lower.9pt\hbox{\cmsss Z\kern-.4em Z}}
{\lower1.2pt\hbox{\cmsss Z\kern-.4em Z}}\else{\cmss Z\kern-.4em Z}\fi}

\def\oneandahalfspace{\baselineskip=\normalbaselineskip \multiply
\baselineskip by 1}

\newcount\equationnumber
\advance\equationnumber by1
\def\ifundefined#1{\expandafter\ifx\csname#1\endcsname\relax}
\def\docref#1{\ifundefined{#1} {\bf ?.?}\message{#1 not yet defined,}
\else \csname#1\endcsname \fi}
\def\autoeqnum{\def\eqlabel##1{\edef##1{\the\equationnumber}}}
\def\no{\eqno(\the\equationnumber){\global\advance\equationnumber by1}}
\newcount\citationnumber
\advance\citationnumber by1
\def\ifundefined#1{\expandafter\ifx\csname#1\endcsname\relax}
\def\cite#1{\ifundefined{#1} {\bf ?.?}\message{#1 not yet defined,}
\else \csname#1\endcsname \fi}
\def\autocite{\def\citelabel##1{\edef##1{\the\citationnumber}\global\advance\citationnumber by1}}

\hsize=6.5truein
\hoffset=.1truein
\vsize=8.9truein
\voffset=.05truein
\parskip=\medskipamount
\twelvepoint           
%\doublespace
\oneandahalfspace
\autocite
\autoeqnum 

% Some macros %

\def\ss{\scriptscriptstyle}
\def\cl{\centerline}

%\line{\hfill ICN-UNAM-97-11}
%\line{\hfill 8th July, 1997}
\vskip 1truein
\cl{\bigg {SYMMETRY BREAKING AND ADAPTATION:}}
\vskip 0.3truein
\cl{\bigg {EVIDENCE FROM A ``TOY MODEL'' OF A VIRUS}}
\vskip 0.3truein
\cl{\bgg J. Mora Vargas}
\vskip\baselineskip
\cl{\it Facultad de Ingenier\'\i a, UNAM}
\cl{\it Circuito Exterior, C.U. }
\cl{\it M\'exico D.F. 04510.}
\vskip\baselineskip
\cl{\bgg  H. Waelbroeck and C.R. Stephens}
\vskip\baselineskip
\cl{\it Instituto de Ciencias Nucleares, UNAM,}
\cl{\it Circuito Exterior, A.Postal 70-543,}
\cl{\it M\'exico D.F. 04510.}
\vskip 1truein

\noindent{\bf Abstract:}\ \ 
We argue that the phenomenon of symmetry breaking in genetics can enhance 
the adaptability of a species to changes in the environment. In the case 
of a virus, the claim is that the codon bias in the neutralization 
epitope improves the virus' ability to generate mutants 
that evade the induced immune response. We support our claim with a 
simple ``toy model'' of a viral epitope evolving in competition
with the immune system. The effective selective advantage of a higher 
mutability leads to a dominance of codons that favour non-synonymous 
mutations. The results in this paper suggest the possibility of emergence
of an algorithmic language in more complicated systems.
\vfill \eject

\line{\bgg 1. Introduction\hfill}

How does evolution work? The XIX'th century witnessed a competition 
between two points of view, best described through the famous example 
of the giraffe's neck. The first point of view, due to Lamarck, 
was that during a giraffe's lifespan the need to constantly stretch 
to reach leaves high up in the trees was somehow understood by the 
reproductory system which would have the giraffes produce offspring with 
longer necks than their parents. The second point of view, due to 
Darwin, was that the offspring are born with a random distribution 
of neck sizes, with a mean value identical to that of their parents. 
The principle of natural selection is then assumed to favour longer 
neck sizes through the differential in reproduction rates. 

 Lamarck's ideas have been mostly relegated to the role of historical 
anecdote with the improved understanding of the biomolecular mechanisms
involved in the manipulation of genetic information. The ``central dogma'',
which is almost always true as far as we know today (Lewin, 1995),
holds that information flows only one way: from chromosome to protein, but 
not the other way around. Yet the discussion has never completely died, 
mainly because the claim that mutations are strictly random is difficult to
reconcile with the observed efficiency of evolution. Some of the main objections are: 

\noindent 1. Simultaneous changes of several apparently independent phenotypic 
traits, which are required to explain many ``large'' mutations (e.g. reptiles 
and birds), seem too improbable to occur without some form of organization.

\noindent 2. The efficiency with which species adapt to 
changes in the environment suggests that there should be a mechanism 
for environment-feedback which favours useful mutations over random ones. 

 In this article, we will show that the environment (e.g. the immune system)
can organise the search for new genetic solutions {\it within the context of
random mutations of the chromosome}. The essential idea is that random
mutations of the {\it genotype} (genetic makeup of an organism) 
produce ``organised'' mutations of the {\it phenotype} (shape and 
function of an organism). In other words, mutations at the 
phenotypic level can be ``guided'' without the need to appeal to a mechanism 
which would violate the central dogma or causality. What we mean by guided 
can be understood once again with the example of the giraffe; here the claim  
would be that the distribution of giraffe offspring is biased toward longer necks. 
To understand how this can occur it is important to recall how 
chromosomes encode genetic information.

 The genetic information is encoded in ribonucleic acid (RNA) and
deoxyribonucleic acid (DNA), these molecules being polymers of four types of
nucleotides (monomers). A group of three nucleotides forms an aminoacid
residue, called a ``codon''. There are 64 possible codons ($4^3$), which
encode a total of $20$ possible aminoacids and a STOP sequence. The code is 
almost universal but variations exist (Lewin, 1995; Trifonov, 1987). 
Since there are $64$ codons and only $20$ amino acids, several
codons can code for the same aminoacid, for example ATT, ATC and ATA all 
represent Isoleucine. In general, the genetic code can be represented most
conveniently by placing the $64$ codons on the 
vertices of a six-dimensional hypercube (Jimen\'ez-Monta\~no, 1996). 

 The translation process produces a chain of amino acids which eventually 
folds into a particular shape that characterises a functional object 
at the microscopic level (protein or enzyme). Here again there is a great 
redundancy. There are 
many possible amino acid substitutions along the polymer that would 
leave the resulting protein unaltered. Finally, the chemical 
interactions and catalytic properties of these products underlie a 
complex biochemical ``computer'' which regulates the organism's development 
and eventually determines the phenotype, e.g. the giraffe's neck size. 
The entire process, which begins with $t-RNA$ molecules binding to 
$m-RNA$ anticodons in rybozymes to synthesise proteins and culminates 
with the macroscopic shape and function of each organ has been described 
as a process of ``percolation through scales'' (Conrad, 1996).

 The importance of the distinction between phenotype and genotype in 
Darwinian evolution has been stressed previously by other authors
(e.g. Gatlin, 1972; Ratner, 1983). Likewise, the concept 
of self-organization in evolution is not new (e.g. Kauffman, 1993). One might 
summarise this paper by saying that it is an attempt to apply the general 
ideas of self-organization to the genotype-phenotype map.
The point which we stress here is that in general this map is non-injective 
there existing a great number 
of ``synonymous'' genotypes for each phenotype. In nature, not all 
synonyms are observed and those that are come in different proportions. It is 
known from the theory of branching processes (Taib, 1992) and the Neutral Theory 
of molecular evolution (Kimura, 1983) that symmetry-breaking occurs spontaneously in a 
finite breeding pool. In this paper we will show that {\it induced 
symmetry breaking } also occurs, due to the violation of the synonym 
symmetry by genetic operators.

 A simple example shows the non-equivalence of different synonyms under the
mutation operator: Consider the synonyms {\it dead} and{\it \ defunct}. A 
point mutation is defined to be the change of single letter. The word 
{\it dead} can mutate to {\it deed}, {\it bead}, {\it lead}, {\it deaf}, 
{\it dean}, {\it dear}, {\it read} or {\it deal}, but is difficult to 
generate a meaningful word by mutating the word {\it defunct}. As we will see
synonymous codons can likewise differ in their mutability. 
Of course it would be very convenient to choose {\it among the synonyms} 
those that have the best mutation targets, thereby preparing the organism 
for mutation to useful products. The question is: how can this occur 
without violating causality or the central dogma? We answer this question 
in the following paragraphs. 

The growth of an individual over many generations  
will take into account not only its own selective advantage, 
but also its ability to produce well-adapted offspring which can in
turn produce well-adapted offspring, etc. One can define an average {\it effective 
fitness} (Stephens and Waelbroeck 1996, 1997) of an 
individual as its growth rate over many generations.
This effective fitness function does not respect the synonym symmetry and
gives rise to a selective pressure which enhances the production of
potentially successful mutants by selecting, among the synonyms, those that
have a higher probability to generate well-adapted offspring. This proposal
implies the existence of a mechanism for environmental feedback in 
the genetic search. Since symmetry
breaking is due in part to selective pressures, some of these by the
environment, it is reasonable to expect that mutant phenotypes are 
produced in an organised manner. What we are saying is in no way in 
contradiction with the central dogma (like Lamarckism): Information from 
the environment is incorporated not through a single individual but 
indirectly through the symmetry breaking of the gene pool. There is
also no violation of causality. The symmetry breaking supports the right 
mutability strategies only to the extent that the mutability
strategies that were useful in the past will continue to be useful in
the future. 

 According to this point of view, it is not the giraffe's neck size that is
being selected but the tendency to produce mutant offspring with longer necks
than their parents. We stress once again that this does not require
that mutations of the {\it chromosomes} be organised, as a modern 
interpretation of Lamarkism would suggest. 
This is important because transcription errors are known to occur 
without premeditation and when directionality is observed there
is no evidence that this directionality is related to environmental 
constraints. The information which allows the mutations to be organised 
{\it at the phenotypic level} is encoded in the distribution of synonyms,
and expressed through the action of a genetic ``interpreter'', i.e. the 
genotype-phenotype map. Due to the 
complexity of the genetic interpreter, it is not difficult to imagine that
two synonymous chromosomes for identical giraffes could differ in the 
sense that one would mutate more easily to a longer neck and the other 
to a shorter neck length. For the sake of illustration we can mention one 
possible mechanism for this sort of predisposition in mutability: Some genes 
are repeated in several sites along the chromosome but without the
required promoter, or with some other genetic ``mistake'' which impedes
its expression. Such redundant copies of a gene for an enzyme which would
stimulate the growth of longer necks can be activated with a simple 
mutation. This is just one of many examples of how the choice of a synonym 
can favour mutability. Of course this is a trivial example and almost certainly 
not the correct explanation of the giraffe phenomenon, but it serves its 
purpose as a feasability proof. 

 The actual mechanism whereby the symmetry 
breaking and ensuing emergence of a genetic language (Popov, 1996) can 
organise mutations of the phenotype is likely to be at least as 
complicated as the genetic interpreter itself, in part because there 
is no evidence that short, compact ``words'' should be more important
than non-local forms of information storage (Stephens and Waelbroeck 
1996, 1997). This suggests using 
simple toy models to try to understand the phenomenon as a 
prerequisite to any serious attempt to explain or demonstrate the 
existence of analogous processes in natural evolution. 

 An example of a simple organism where a mutation strategy can be 
selected and where evidence of a language exists is the evolution of 
a viral neutralization epitope {\it in vivo} (Burgos,1996; Vera and 
Waelbroeck, 1996). In this case in order to evade the induced 
immune response the best evolutionary strategy is to be as mutable as possible. 
Since this strategy is valid at all times, it is a situation where the 
symmetry breaking, which reflects the selective pressure to mutate in the 
past, can also be expected to function in the future, by again 
enhancing the ability of the epitope to mutate. In a separate article
an analysis of the coding of the {\it env} proteins was carried out 
and provided evidence in favour of our proposal. However, in that case 
several other selective factors are competing with  
the need to mutate; for example m-RNA secondary structure, enzyme and
t-RNA availabilities in the infected cells, and secondary structure 
constraints all play some role. It is also not known precisely what 
segment(s) of the V3 loop region or other hypervariable regions is (are)
recognised as a neutralization epitope by inmunoglobulins. 

 Our purpose in this paper is to provide cleaner evidence for our proposed 
mechanism (whereby adaptation is enhanced by symmetry breaking), using 
a simple toy model where the {\it only} selective factor is the need
to adapt to the immune response. 

\vskip 0.3truein

\line{\bgg 2. Model and Results\hfill}

In this section we will present the ``toy'' model with which we will 
illustrate our ideas. The model consists of a virus, of which we will
consider the evolution of one ``epitope'' represented by six ``aminoacids''.
There are three possible amino acids ($a$,$b$,$c$) at each position represented by 
three-bit ``codons''. The genotype then is an $18$-bit ``chromosome''. 
The possible values for each ``nucleotide'' are $0$ or $1$, i.e. there are 
two bases, hence there are $8$ possible codons. The total 
number of possible genotypes is $2^{\ss 18}= 262,144$. 
The phenotype is a six-letter word, e.g. $aabacb$, each word representing a
different ``virus''. The total number of possible phenotypes is $3^{\ss 6}$=729. 

As there are eight possible codons and only three 
aminoacids the genotype-phenotype mapping will be degenerate.
This is manifest in the difference between the total number of genotypes and the
total number of phenotypes. If we think of the 
genotype-phenotype mapping, $\phi$, as an ``interpreter'' then $\phi$ is non-injective.
We specify the interpreter map $\phi$ at the level of codon and aminoacid. Specifically: 
$\phi(000)=\phi(001)=\phi(010)=a$, $\phi(011)=\phi(100)=\phi(101)=b$ and
$\phi(110)=\phi(111)=c$.

Each codon has a different proper mutability, where by proper
mutability we mean the number of different aminoacids reached
by any point mutation in a particular codon. The relation between
codon, aminoacid and proper mutability is shown below 
$$\vbox{\settabs 4 \columns
\+Aminoacid&Codon&Proper mutability& Aminoacid reached\cr 
\+\ \ \ \ a&\ 000&\ \ \ \ \ 1&b(100)\cr
\+\ \ \ \ a&\ 001&\ \ \ \ \ 2&b(101), b(011)\cr  
\+\ \ \ \ a&\ 010&\ \ \ \ \ 2&b(011), c(110)\cr  
\+\ \ \ \ b&\ 011&\ \ \ \ \ 3&a(010), b(001), c(111)\cr 
\+\ \ \ \ b&\ 100&\ \ \ \ \ 2&a(000), b(101)\cr 
\+\ \ \ \ b&\ 101&\ \ \ \ \ 2&a(001), c(111)\cr
\+\ \ \ \ c&\ 110&\ \ \ \ \ 2&a(010), b(100)\cr  
\+\ \ \ \ c&\ 111&\ \ \ \ \ 2&b(011), b(101)\cr}$$ 

\centerline {Table 1}
As can be seen, for most codons the proper mutability is 
$2$. The codon $000$ however has a proper mutability $1$ whilst 
$011$ is the most mutable codon with a proper mutability of $3$.  
The proper mutability of the entire chromosome is by definition the sum of the 
proper mutabilities of its six constituent codons. In an initial 
random population it is $12$ on average and ranges from $6$ 
(when all codons are $000$) to $18$ (when all codons are $011$). 

The fitness, $f_i(n)$, of the ith virus at generation $n$ is a measure of how 
well the virus evades the immune system. Fitness, however, is associated
with a phenotype not a genotype, hence to calculate the fitness the genotype
must be translated to a phenotype. For example
\vskip 0.1truein
\halign{
# & # & # & # & #\cr
\hskip 2 truecm &\ \ \ \ Genotype \hskip 2.5 truecm &Genotype number\ \ \  &Phenotype \ \ \ &Phenotype number\cr
&$000010111110110101$&$\ \ \ \ 12213$&$\ aacccb$&$\ \ \ \ \ 79$\cr  
&$101111000001101001$&$\ \ \ \ 192617$&$\ bcaaba$&$\ \ \ \ \ 408$\cr}
\noindent The fitness of a particular viral strain, $i$, obeys the equation
$$f_i(n+1)=f_i(n)-\gamma V_i(n),\no$$
where $V_i(n)$ is the amount of virus $i$ present at time $n$ and
$\gamma$ is a constant which represents the immune system's success 
in recognizing a viral strain and responding by creating T-cells capable of
attacking this particular strain. The decrease in fitness of a virus in proportion
to its abundance is a representation of the action of 
macrophages which consume viruses and activate T-cells which recognise 
the specific neutralization epitope of the consumed virus.

For the evolution of $V_i(n)$ we used a 
selection operator that replaced $V_i(n)$ by the integer part of  
$f_i(n)V_i(n)R\zeta$, where we introduce $R$ as a reproduction parameter and
$\zeta $ is a random number uniformly distributed in the unit interval. After selection
we implement mutation with probability $\mu$ per bit.
A mutation, which acts on genotypes in a completely random 
and symmetrical fashion, flips the gene value: if it was $1$ it will be $0$ and
viceversa. The mutation rate is previously determined and is the same for
the entire experiment. 

The system is initialized with a single virus with unit fitness. 
A ``generation'' consists of: evaluation of the fitness of each virus; 
reproduction of each virus and finally mutation of the selected chromosomes 
As the evolution proceeds 
the initial fitness of any mutant virus not previously present in the 
population is set to $1$. When the fitness reaches $0$ this signifies that the
virus cannot survive the induced immune response (T cells) and so it 
disappears completely from the population. It can also disappear prior 
to this due to the random sampling effect inherent in a finite breeding pool. 
A virus which has been successfully eliminated in one generation 
can of course reappear later through a mutation of
a different strain. When this happens it reappears with a fitness 
value equal to $1$. 

The model above tries to represent faithfully the essence of a 
phenomenon which we claim is found in nature (Waelbroeck, 1997). 
A ``virus'', which we 
have identified with the phenotype in our model, consists of a chain 
of just $6$ aminoacids which is roughly the size of the neutralization 
epitope. From the point of view of the immune system this is what 
characterises the virus. 

When the total amount of virus exceeds a given number, $M$, the
infected person dies and the evolution stops. If the total amount
of virus does not exceed $M$ within $G$ generations then the 
patient has survived. The particular values of the parameters chosen
were: $\mu=0.001$, $\gamma =0.001$, $M=1000$ and $G=3000$.
The long ``latency'' period ($G=3000$) is achieved by ``weakening"  the patient 
when his virus count reaches a low value and ``treating" him 
when it is excessively large, to avoid both recovery and death of 
the patient. More precisely, if the total virus count is 
over $600$ the reproduction parameter, $R$, is set to $3$ and if it is below $5$ 
the parameter is set to $8$. For intermediate values we set $R=4$.
With these choices some runs with long incubation periods are observed; the
code was designed to only retain information about the infections that lasted 
the full 3000 generations.

The fact that $\gamma=0.001$
implies that if $1000$ copies of a particular virus have been detected
over a period of time, the induced immune response is perfect and 
the virus is necessarily eliminated as its fitness goes to zero.
With the value $4$ for the reproduction factor, the expected time for 
the immune response to eliminate an infection by a particular strain 
is about ten generations; if infection has a duration of $3000$ generations 
this implies that we will be observing about $300$ mutations.

The proper mutability was averaged during the $3000$ generations of 
each experiment. In Figure 1 the results of $1000$ experiments 
are represented. The graph shows that all the experiments had an average 
proper mutability greater than $12$. This is evidence of
environmental feedback wherein the environment is represented by the immune 
system in the way it decreases the fitness of a viral strain as 
it becomes recognised. Since the virus is constantly forced to mutate 
to new forms, which by definition start with a high fitness, to avoid extermination, 
a strategy is selected whereby codons which mutate more frequently 
to non-synonymous targets are preferred. As mentioned above this implies using more
frequently the codon $011$, which has a proper mutability of $3$, and less 
frequently the codon $000$ which mutates mostly to other synonyms.

In Fig. 2 we show the average virus diversity during the infection 
for each of the $1000$ experiments. In all cases the ratio of virus 
strains present to the total number possible is less than $15\%$.
Since mutation acts on every gene, if the population grows, the number of new
viruses grows as well, so a high diversity is related to an increase in the
population.

 Figures 3 to 6 we show two ``experiments" in extenso. In Figure 3 one can see 
that the patient almost eliminates the virus completely in the first generations; the 
proper mutability eventually increases to an exceptionally high value, above 13 (Figure 4).
The run described in Figures 5,6 is more ``generic''. One can note the fluctuations 
of the total infection during the process, where the total viral population ranges from
500 to 750. Figs. 4 and 6 show that the proper mutability is above 12 for almost 
all generations, thereby confirming our hypothesis about symmetry breaking.

\vskip 0.3truein
 
\line{\bgg 4. Discussion and Conclusions\hfill}

Our model shows the spontaneous emergence of structure in the production
of mutants in spite of the purely random and non-directional 
nature of point mutations at the genotypic level. 
This reflects a self-organization process which improves the virus'
ability to adapt to changes in its environment. 
The proper mutability average for virus with a long lifetime {\it in vivo} 
is above the average value, reflecting the dominance of the more mutable codons 
in the chromosome. Thus, we have demonstrated that a virus can ``organise'' its 
mutations in order to avoid the immune system response through 
the choice of non-synonymous mutations over neutral ones.

 From all possible epitopes that could be represented in our model, only a 
small proportion, approximately $15 \%$ , are actually present at any time 
during the infection. The only selective difference between one epitope 
and another is the extent to which the immune system has learned to recognise 
it. Thus, any mutation to one of the remaining $85\%$ of as yet unrecognised 
epitopes is selectively favoured. The effective selective advantage in the
long term of a high proper mutability leads to the dominance of codons 
that favour non-synonymous mutations.

The induced symmetry breaking which results in the enhanced usage of mutable codons
is closely related to the non-injective property of the interpreter which 
carries out the translation process. This implies that synonyms can exist 
which differ from one another only by the action of the genetic operators; 
this action induces a hierarchy among synonyms, which leads to the symmetry
breaking. For the case treated in this paper, the relevant genetic operator is
point mutation and the hierarchy is due to
proper mutability. The interpreter used in our experiment was trivial in that
its only role was to establish three possible hierarchical states 
depending on proper mutability, with mutability values $1$, $2$ and $3$. 
Since the codon bias occurred within this simple model one is led to the
conclusion that the property of mutability itself is being 
selected. The interpreter is independent of
the fitness landscape which depends only on the phenotype.

 Of course, one may take the point of view that there is no fundamental 
difference between a non-injective fitness function on genotypes (i.e., 
direct encoding with a ``symmetrical'' fitness landscape) and 
a non-injective interpreter (indirect encoding). However, by following this
point of view one misses the point, which is that by focusing on 
the role of the interpreter we can better understand how evolution works. 
In this case, identifying the interpreter allows us to understand that the 
origin of the symmetry of the induced fitness function on genotypes is 
that different codons can code for the same amino acid but differ from 
one another in mutability. This allows the system to better face 
the changes needed for adaptation. 

 In conclusion, our model proves that symmetry breaking can enhance the 
adaptability of a species to changes in the environment. The production
of phenotypic mutations is not only a reflection of the random and 
non-directional nature of point mutations, but of the spontaneous
emergence of structure in the gene pool through symmetry breaking. 

 For this mechanism to be useful towards understanding the 
self-organization of phenotypic evolution in more complex organisms 
the concept of synonym must be generalised beyond the simple 
codon-aminoacid redundancy considered here. The
chromosome does not encode directly the size and shape of various parts of
an organism, but rather the {\it interpreter}, in this case embodied by biochemical 
processes in living cells (and amongst them), translates the genotype 
into a phenotype. In this translation there are many possible sources 
of redundancy, the codon bias being only a relatively insignificant 
example. There are more subtle forms of synonyms, involving issues ranging 
from protein secondary structure to the machinery of gene regulation, 
for which symmetry breaking can be related to the emergence of an {\it 
algorithmic language.}

 Considering the chromosome (genotype) as an algorithm, the interpreter is
the ``computer'' which executes the algorithm and the phenotype is the
solution. In this sense, the breaking of symmetry is related to the
selection of a language, where ``words'' or ``grammatical rules'' are
selected in order to facilitate the search for well-adapted offspring, 
i.e. successful mutants. The identification of such subunits of 
genetic information (Schmitt, 1996) which facilitate the search for mutant phenotypes is
related to the standard problem of finding an approximate decomposition
of an optimization problem into smaller subproblems. The condition for 
such a strategy to succeed is that when the solutions to the subproblems 
are reached then a good approximation to the global solution is reached
as well. This requires that the fitness landscape should have a certain 
amount of structure; by unravelling this structure the emergent language 
results in an effective smoothing of the induced landscape on the space 
of genotypes. By ``effective smoothing'' we mean the population-dependent property that 
mutations at the level of the genotype have better mutation targets on 
average than in a random population. This implies first of all a solution 
of the brittleness problem, since the first task is that mutant algorithms
be meaningful, and secondly an enhanced ability to produce genetic 
improvements (better algorithms). For the proposed mechanism to work it
is necessary that the landscape be sufficiently correlated, and 
that the interpreter be well adjusted to the structure of the problem. An
example would be the Kauffman's {\it Nk} landscapes for {\it k} $\ll$ {\it N,} 
together with his model of cellular automata for gene regulation. Another
example (Angeles et al 1997) is the cell division interpreter in Kitano's
neurogenetic model (Kitano 1990, 1994).

 In the case of complex interpreters that lead to algorithmic languages , 
in order that the symmetry breaking which necessarily reflects only 
{\it past} adaptation pressures should favour the search of {\it future} 
solutions, the evolution of the landscape must respect certain rules. 
Namely, the decomposition of the optimization problem into subproblems
must be independent of time, so that the algorithmic language which has 
been successful in past should continue to be useful in future.
This is the requirement of {\it structural decomposition stability}: 
The landscape evolution must preserve the structural decomposition 
of the adaptation problem.

 One might conjecture that extinctions are related to a violation of
structural decomposition stability. For instance, the algorithmic language
guiding the search of new dinosaur species would presumably have been
incapable of producing viable solutions in the environment which is assumed
to have provoke their demise.

 The symmetry breaking which we observe in these experiments support these 
ideas by suggesting that with a less trivial interpreter one might witness 
the emergence of an ``algorithmic language'' tuned to the interpreter. 

 We are currently analyzing several Genetic Algorithm models to this effect,
(Holland, 1975; Goldberg, 1989),
using certain classes of controlled rugged landscapes that are more
realistic from a biological point of view (Kauffman, 1989,1990,1993).
A related challenge is to exploit the emergence of a language to assist 
in the design of a new generation of genetic algorithms as an improved 
general purpose optimization method. A key for success in this direction is 
the codification method: The interpreter should have the sufficient 
flexibility to be able to solve the decomposition problem, but not so much
flexibility that it could solve any possible problem, since in that case the
search space for the desired algorithmic language would be far too large. 
Another application of the language emergence is the development of an GA 
to perform complex computational tasks (Crutchfield 1994), such as combinatorics. 
Interesting applications may also follow in adaptive systems modelling 
where adaptability is an important property, for example
the forecasting problem in financial markets. 

\vskip 0.1truein
\noindent {\bf Acknowledgements} We are grateful to 
the entire Complex Systems group under {\it NNCP} 
(http://luthien.nuclecu.unam.mx/~nncp) for maintaining a 
stimulating research atmosphere. This work is supported in part by the 
DGAPA-UNAM grant IN105197. One of us (JM) would also like to 
acknowledge support from CONACyT through a doctoral scholarship.

\vskip 0.2truein
\centerline{\bgg References}

\noindent Angeles O, Waelbroeck H and Stephens, C R (1997) Emergence of Algorithmic Language in Genetic Systems. National University of Mexico Preprint ICN-UNAM-97-08, (adap-org/9708001) submitted to BioSystems.

\noindent Blitz D (1992) Emergent Evolution: Qualitative Novelty and the Levels 
of Reality.  Kluwer, Dordrecht.

\noindent Burgos J D and Moreno-Tovar P (1996) Zipf-scaling Behavior in 
the Immune System. BioSystems {\bf 39}: 227-332

\noindent Conrad M (1996) Cross-scale Information Processing in Evolution, 
Development and Intelligence. BioSystems {\bf 38}: 97-109

\noindent Crutchfield J P (1994) The Calculi of Emergence: Computation, Dynamics
and Induction. Physica {\bf D 75}: 11-54

\noindent Gatlin L L (1972) Information Theory and the Living System. 
Columbia University, New York Press, N.Y.

\noindent Goldberg D E  (1989) Genetic Algorithms in Search, 
Optimization and Machine Learning. Addison Wesley, Reading, MA.

\noindent Holland J H (1975) Adaptation in Natural and Artificial Systems. 
MIT Press, Cambridge, MA. 

\noindent Jimen\'ez-Monta\~no M A {\it et al} (1996) The Hypercube Structure 
of the Genetic Code Explains Conservative and Non-conservative Aminoacid
Substitutions {\it in vivo} and {\it in vitro}. BioSystems {\bf 39}: 117-125

\noindent Kauffman S A (1989) Adaptation on rugged fitness landscapes, in: 
Lectures in the Sciences of Complexity. D. Stein (ed.) (Reading, 
MA, Addison-Wesley) pp. 527-618

\noindent Kauffman S A (1990) Dyanamics of Evolution. Lecture at the Workshop 
Complex Dynamics and Biological Evolution. Hindsgavl, Physica {\bf D42}:135

\noindent Kauffman S A (1993) The Origins of Order. Self-Organization and 
selection in Evolution. Oxford University Press, New York, Oxford

\noindent Kimura M (1983) The Neutral Theory of Molecular
Evolution. Cambridge University Press, Cambridge

\noindent Kitano H (1990) Designing Neural Networks Using Genetic Algorithms
with a Graph Generation System. Complex Syst. {\bf 4}: 461-476

\noindent Kitano H (1994) Neurogenetic Learning: an Integrated Method of Designing and 
Training Neural Networks Using Genetic Algorithms. Physica D {\bf 75}: 225-238

\noindent Lewin B (1995) Genes V. Oxford University Press, Oxford 

\noindent Popov O {\it et al} (1996)  Linguistic Complexity of Protein Sequences 
as Compared to Texts of Human Languages. BioSystems {\bf 38}: 65-74

\noindent Ratner V A (1983) Molecular Genetics, Principles and Mechanism (in 
Russian). Nauka Novosibirsk

\noindent Schmitt A O {\it et al} (1996) The Modular Structure of Informational 
Sequences. Biosystems {\bf 37}: 199-210

\noindent Stephens C R and Waelbroeck H (1996) Analysis of the Effective Degrees 
of Freedom in Genetic Algorithms. National University of Mexico Preprint 
ICN-UNAM-96-08 (adap-org/9611005), submitted to Phys. Rev. E

\noindent Stephens C R and Waelbroeck H (1997) Effective Degrees 
of Freedom of Genetic Algorithms and the Block Hypothesis. National University 
of Mexico Preprint ICN-UNAM-96-08 (adap-org/0797), to appear in the Proceedings
of the Sixth International Conference on Genetic Algorithms, (Morgan Kaufman 1997).

\noindent Taib Z (1992) Branching Processes and Neutral Evolution. Springer 
Verlag, Berlin

\noindent Trifonov E N and Brendel V (1987) A Dictionary of Genetic 
Codes. VCH, Weinheim

\noindent Vera S and Waelbroeck H (1996) Symmetry Breaking 
and Adaptation: the Genetic Code of Retroviral {\it env} Proteins. 
National University of Mexico Preprint ICN-UNAM-96-09 (adap-org/9610001)

\noindent Waelbroeck H (1997) Codon Bias and Mutability in HIV Sequences. National 
University of Mexico Preprint ICN-UNAM-97-09 (adap-org/9708003).

\vfill\eject

\end